%% file: diagonalEnsemble.tex
\documentclass[aps,prb,twocolumn,superscriptaddress,amsmath,amssymb,preprintnumbers]{revtex4-2}

\usepackage{epsfig,afterpage}
\usepackage{graphicx}% Include figure files
\usepackage{amsfonts}
\usepackage{amssymb}
\usepackage{hyperref}
\usepackage{nicefrac}
\usepackage{indentfirst}
\usepackage{amsmath,amsthm}
\usepackage{dsfont}
\usepackage{epsfig}
\usepackage{subfigure}
\usepackage{ulem}
\usepackage{multirow}
\usepackage{pstricks}

\usepackage{wasysym} % smiley
\usepackage{bbold} % Id

\def\beq{\begin{equation}}
\def\eeq{\end{equation}}
\def\bal{\begin{align}}
\def\eal{\end{align}}
\def\bea{\begin{eqnarray}}
\def\eea{\end{eqnarray}}
\def\bq{\begin{quote}}
\def\eq{\end{quote}}
\def\ben{\begin{enumerate}}
\def\een{\end{enumerate}}
\def\bit{\begin{itemize}}
\def\eit{\end{itemize}}

\newcommand{\ket}[1]{|#1\rangle}
\newcommand{\bra}[1]{\langle#1|}
\newcommand{\ev}[1]{\langle #1 \rangle}
\def\Id{\mathbb{1}}
\def\tr{\mathrm{tr}}
\def\Hc{H_C}
\def\hatHc{\hat{H}_C}

\begin{document}
	
	\title{Approximating the long time average of the density operator: Diagonal ensemble}
	
	\author{Asl{\i} \surname{\c{C}akan}}
	\affiliation{Max-Planck-Institut f\"ur Quantenoptik, Hans-Kopfermann-Str. 1, 85748 Garching, Germany} \affiliation{Munich Center for Quantum Science and Technology (MCQST), Schellingstr. 4, D-80799 Munich, Germany}
	\author{J. Ignacio \surname{Cirac}}
	\affiliation{Max-Planck-Institut f\"ur Quantenoptik, Hans-Kopfermann-Str. 1, 85748 Garching, Germany} \affiliation{Munich Center for Quantum Science and Technology (MCQST), Schellingstr. 4, D-80799 Munich, Germany}
	\author{Mari Carmen \surname{Ba\~nuls}}%\email{banulsm@mpq.mpg.de} 
	\affiliation{Max-Planck-Institut f\"ur Quantenoptik, Hans-Kopfermann-Str. 1, 85748 Garching, Germany} \affiliation{Munich Center for Quantum Science and Technology (MCQST), Schellingstr. 4, D-80799 Munich, Germany}
	
	%\date{\today}
	%\tableofcontents
	
	\makeatletter
	\let\toc@pre\relax
	\let\toc@post\relax
	\makeatother 
	\newpage
	
	\begin{abstract}
	For an isolated generic quantum system out of equilibrium, the long time average of observables is given by the diagonal ensemble, i.e. the mixed state with the same probability for energy eigenstates as the initial state but without coherences between different energies. In this work we present a method to approximate the diagonal ensemble using tensor networks. Instead of simulating the real time evolution, we adapt a filtering scheme introduced earlier in [Phys. Rev. B 101, 144305 (2020)] to this problem. We analyze the performance of the method on a non-integrable spin chain, for which we
	observe that local observables converge towards thermal values polynomially with the inverse width of the filter.
	\end{abstract}
	
	\maketitle
	
	\section{Introduction}
    When an isolated quantum system is initialized in a pure state out of equilibrium, the unitary character of the evolution ensures that the state remains pure at any later times. However, if observations are restricted to a subsystem, thermalization may occur, that is, the rest of the system can act as a bath for the observed region~\cite{Deutsch1991, Rigol2008}.
	More explicitly, if expectation values reach and remain close to a certain value for an extended period of time, one talks about equilibration~\cite{Srednicki_1999,Masanes2013,Gogolin_2016}. And thermalization occurs if those values correspond to the expectation values at the thermal equilibrium state consistent with the energy of the system~\cite{Srednicki1994,Deutsch1991, Berges_2001,Rigol2008}.
	
For a generic Hamiltonian with non-degenerate spectrum, the long-time limit of time-averaged observables corresponds to the expectation value in the diagonal ensemble~\cite{Rigol2007}. This mixed state, diagonal in the energy eigenbasis, can be seen as the average of the density operator of the system at all times. To decide whether the system can thermalize it is thus enough to compare the expectation values in the diagonal ensemble to those in thermal equilibrium at the same energy.
But while the thermal state of a local Hamiltonian can be efficiently approximated using tensor networks~\cite{Hastings2006,Molnar2015,Kuwahara2020}, simulating the out-of-equilibrium dynamics, and thus directly constructing the diagonal ensemble, is a much harder problem~\cite{osborne2006,Schuch2008}.

Generally speaking, integrable systems, due to their extensive number of conserved local quantities, do not thermalize but are instead argued to relax or equilibrate to the so-called generalized Gibbs ensemble~\cite{Rigol2007,Rigol2008,Cramer2008,DAlessio2016}, compatible with all the constraints. In contrast, non-integrable systems are typically expected to thermalize~\cite{Kollath2007, Manmane2007,Flesch2008,Moeckel2008,Rigol2008,Rigol2009,Gogolin_2016}. 
It is thus especially interesting to identify non-integrable systems that fail to do so, as the current interest in systems with many body localization~\cite{Cassidy2009,Santos2010,Abanin2019}, quantum scars~\cite{Turner2018} or disorder-free localization~\cite{PAPIC2015,Smith2017,Schulz2019} makes evident. Nevertheless, the (absence of) thermalization of non-integrable systems is hard to determine, since the applicability of analytical tools for such models is limited, and numerical simulations of out of equilibrium dynamics are restricted to small systems or short times.

In this paper, we present an alternative method to approximate the diagonal ensemble without resourcing to the explicit simulation of the dynamics. We make use of a recently introduced filtering procedure~\cite{banuls2019entanglement}, devised to prepare pure states with reduced energy variance, and show how it can be adapted to filter out the off-diagonal components of a density operator with respect to the energy basis.

More concretely, we apply to the initial density matrix a 
Gaussian operator that filters out large eigenvalues of the Hamiltonian commutator. In the limit of vanishing width of the Gaussian, the result will converge to the diagonal ensemble, in the most generic case, when there are no degeneracies in the spectrum. Notice that if there were degenerate energy levels, the procedure would leave untouched the coherences in the corresponding energy subspace, and thus would still lead to the correct limit of the time-averaged density operator. As described in \cite{banuls2019entanglement}, the filter can be approximated as a sum of Chebyshev polynomials, and its application to an initial vector can be numerically simulated using matrix product states~\cite{Verstraete2008,SCHOLLWOCK2011} (MPS) methods, at least for moderate widths. Here we carry out these simulations for a non-integrable spin chain and investigate how the values of local observables converge towards the thermal equilibrium.
	
The rest of the paper is organized as follows. In section~\ref{Filtering} we review the filtering procedure and its application to the problem of the diagonal ensemble. We also discuss some  properties of this specific application. Section~\ref{Methods} describes the main elements of our numerical simulations. Our results are shown in  section~\ref{Results}, where we discuss how the application of the approximate filter to this problem resembles and differs that of reducing the energy variance, and analyze the convergence of local observables to their thermal values. Finally, in Section~\ref{Discussion} we summarize our findings and discuss potential extensions of our work.

\section{Filtering the diagonal ensemble}\label{Filtering}

Let us consider a system of size $N$ governed by a (local) Hamiltonian $H$, and a pure initial state, which can be written in the energy eigenbasis as $\ket{\Psi_0}=\sum_n c_n \ket{E_n}$, with the normalization condition $\sum_n |c_n|^2=1$.
We are interested in the long time average properties of the evolved state, i.e. given any physical observable $O=\sum_{n,m} O_{nm} \ket{E_n}\bra{E_m}$,
we want to compute
\begin{align}
\lim_{T\to \infty} \frac{1}{T}\int dt \bra{\Psi(t)} O \ket{\Psi(t)}&=
\sum_{n} |c_n|^2 O_{nn}=\tr{\left [\rho_D(\Psi_0) O \right]}
\end{align}
where the first equality holds under the generic condition, which we assume in the following, that the spectrum is non-degenerate~\footnote{If this condition is not fulfilled, $\rho_D$ should be replaced by a block-diagonal operator, where each block corresponds to a different energy subspace, with the same matrix elements as in the initial state.},
and in the second one we have used the definition of the diagonal ensemble
\beq
\rho_D(\Psi_0)=\sum_n |c_n|^2 \ket{E_n}\bra{E_n}.
\label{eq:diagEns}
\eeq
If the system thermalizes, the diagonal expectation value $\ev{O}_D :=\tr \left(\rho_D O\right)$ will be equal to the expectation value in the thermal equilibrium state, $\rho_{th}(\beta)=\nicefrac{e^{-\beta H}}{\tr (e^{-\beta H})}$, that corresponds to the mean energy of the initial state.
Thus, an approximation to the diagonal ensemble would allow us to probe whether a given state thermalizes or not.

In the energy eigenbasis, the density matrix for the initial state can be written as $\rho_0=\sum_{n,m} c_n c_m^* \ket{E_n} \bra{E_m}$.  
Filtering out the off-diagonal matrix elements in this basis will result in the diagonal ensemble \eqref{eq:diagEns}. We thus define an (unnormalized) Gaussian filter which acts on the mixed state as a superoperator
\beq
F_{\sigma}[\rho]:=e^{-{\hatHc^2/2 \sigma^2}}[\rho],
\label{eq:Gaussianfilter}
\eeq
where $\hatHc$ is the commutator with the Hamiltonian, i.e. $\hatHc[\rho]=H\rho-\rho H$. 
Notice that $F_{\sigma}$ is a completely positive trace preserving map, i.e. a quantum channel.
The effect of this filter is to suppress the off-diagonal matrix elements corresponding to pairs of states with different energies. 
As the width $\sigma$ is reduced, and for a generic, non-degenerate, Hamiltonian, the application of the filter will converge to the desired result $$F_{\sigma}[\rho_0] \underset{\sigma \to 0}{\longrightarrow} \rho_D(\Psi_0).$$
Notice that the filter would not affect the density operator components in a degenerate energy subspace. Thus, if the Hamiltonian has degenerate levels, the limit of the procedure is block diagonal, corresponding to the long time limit of the time-average of the evolved state.

Mapping the basis operators to vectors~\cite{Choi_1975} as $\ket{E_n}\bra{E_m}\to \ket{E_n E_m}$, we can write the density matrix as a vector of dimension $2^{2N}$, on which the filter acts as a linear operator, and the
problem becomes formally analogous to the energy filters used in~\cite{banuls2019entanglement,lu2020algorithms,Ge2019,Yang_2020}.

In this representation, the commutator corresponds to the linear operator $\hatHc=H\otimes \Id - \Id \otimes H^T$, which, if $H$ is local, is also a local Hamiltonian with eigenvectors $\ket{E_n E_m}$ and corresponding eigenvalues $E_n-E_m$, for $n,\,m=1\ldots {2^N}$.
We can then apply the filtering procedure for reducing the energy variance from a state with given mean energy described in~\cite{banuls2019entanglement}. 
For a product initial state $\ket{\Psi_0}$, the (vectorized) initial density matrix $\ket{\rho_0}=\ket{\Psi_0}\otimes\ket{\Psi_0}$ is also a product, and the scenario is very similar to the one discussed in that reference.

With respect to the Hamiltonian $\hatHc$, any physical state has mean value $\bra{\rho_0} \hatHc \ket{\rho_0}=\tr \left(\rho_0^{\dagger}\, [H,\rho_0]\right)=0$. 
The filter \eqref{eq:Gaussianfilter} preserves this property of the initial state while it reduces the corresponding (effective energy) variance, $\bra{\rho} \hatHc^2 \ket{\rho}=-\tr \left([H,\rho]^2\right)$, 
which measures precisely the off-diagonal part of the density operator in the energy basis.

\subsection{Chebyshev approximation of the filter} \label{Chebyshev approximation of the filter}

Formally, this filtering procedure is analogous to the one described in~\cite{banuls2019entanglement}, and some of the properties can be directly translated to the current case. In particular, the Gaussian filter $F_{\sigma}$ can be approximated 
by a series of Chebyshev polynomials.

Any piece-wise continuous function $f(x)$ defined in the interval $x\in[-1,1]$ can be approximated by a linear combination of the $M$ lowest-degree Chebyshev polynomials~\cite{KPM}. In particular, the corresponding series for the delta function truncated to order $M$ (and improved using the kernel polynomial method) is known to approximate a Gaussian of width $\sqrt{\pi}/M$. We can thus use such series to order $M \propto N/\sigma$ to approximate the Gaussian filter $F_{\sigma}$. This sum has the form 

	\beq
	{Q}_M:=  \sum_{m=0}^{\lfloor{M/2}\rfloor}(-1)^m \frac{2-\delta_{m0}}{\pi}\gamma_{2m}^{M}T_{2m}(\alpha {\hatHc}),
	\label{eq:chebyfilter}
	\eeq
where $\alpha$ is a rescaling constant to ensure that the spectrum of $\alpha \hatHc$ lies strictly within 
the interval $[-1,1]$. We use $\Hc=\alpha \hatHc$ for the rescaled Hamiltonian commutator at the rest of the paper.
$T_m(x)$ is the $m$-th Chebyshev polynomial of the first kind, defined by the recurrence relations
$T_0(x)=1$, $T_1(x)=x$ and $T_{m+1}(x)=2xT_m(x)-T_{m-1}(x)$,
and $\gamma_m^M$ are the Jackson kernel coefficients~\cite{KPM}, 
	\beq
	\gamma_m^{M}= \frac{(M-m+1)\cos{\frac{\pi m}{M+1}}+ \sin{\frac{\pi m}{M+1}}\cos \frac{\pi}{M+1}}{M+1}.
	\eeq

We will denote the result of applying the series expansion to order $M$ as
\beq
\ket{\rho_M}:={Q}_M \ket{\rho_0}.
\label{eq:rho_M}
\eeq
Notice that this vector has a different normalization than  $\ket{\rho_{\sigma}}$, because the sum in ${Q}_M$ approximates a normalized Gaussian distribution, unlike $F_{\sigma}$ from \eqref{eq:Gaussianfilter}.

The off-diagonal width of the operator $\rho_M$ is determined by the corresponding variance of $\Hc$ as
\beq
\delta^2:=\frac{\bra{\rho_M} \Hc^2\ket{\rho_M}}{\bra{\rho_M} \rho_M\rangle}.
\eeq

\subsection{Properties of  the diagonal filter}
\label{subsec:props}

Notice that the filtering procedure described so far is general, as it does not make any assumption on the spatial dimension of the problem.
In the following we will focus on a one-dimensional problem, for which we can use tensor networks in order to obtain numerical approximations.
As in~\cite{banuls2019entanglement}, we can use matrix product state (MPS) techniques~\cite{Verstraete2008,SCHOLLWOCK2011} to simulate the application of this filter to an initial state. In this way we construct a matrix product operator (MPO)~\cite{Verstraete-GarciaRipoll-Cirac_2004,ZwolakVidal2004,Pirvu_2011} approximation to the filtered ensemble. 
Also here, for large system sizes and narrow filters, the
 required bond dimension for the approximation can be bounded as $D\lesssim c' \sqrt{N} D_0
^{1/\delta}$, where $c'$ and $D_0$ are $O(1)$ constants.
Accordingly, the expression for the entanglement entropy,
\beq
\mathcal{S} \lesssim k/\delta + \log \sqrt{N} +\mathrm{const}
\label{eq:ent}
\eeq
corresponds now to a bound for the operator space entanglement entropy (OSEE)~\cite{Prosen_2007}.

The spectrum of $\Hc$ exhibits however an exponential degeneracy in the subspace of eigenvalue zero, which imposes a significant difference. 
For each eigenstate $\ket{E_n}$ of $H$, $\ket{E_n E_n}$ is eigenstate of $\Hc$ with zero eigenvalue.
Thus, even if the spectrum of $H$ is non-degenerate and even if it fulfills the stronger assumption of non-degenerate gaps,  the ``zero energy'' subspace of $\Hc$ is always exponentially degenerate. 

Hence the target diagonal ensemble states could in principle have arbitrarily small OSEE, even with vanishing width $\sigma$ (an extreme case would be the maximally mixed state, with zero OSSE). 
This is in contrast to the Hamiltonian filtering, where the limit would generically have thermal (i.e. volume law) entanglement.
Even if we expect that the general relations between energy fluctuations and entropy or bond dimension demonstrated in~\cite{banuls2019entanglement} still hold during the main part of the filtering procedure,  eventually, as the width becomes negligible and the procedure converges to the diagonal ensemble, the OSSE can converge to a non-generic value that will depend on the initial state.

The scenario we discuss here also exhibits another fundamental difference regarding physical observables. 
For a local operator $O$, the expectation value is computed as 
$$\frac{\tr \left(O \rho \right)}{\tr \rho}=\frac{\bra{O}\rho \rangle}{\bra{\Id}\rho \rangle},
$$
where $\ket{O}$ and $\ket{\Id}$ are respectively the vectorized observable and identity operators.

As an overlap between two vectors, this is a global quantity, and
no longer local in space.
 Therefore, the considerations in~\cite{banuls2019entanglement} about the minimal entanglement of a subregion required for local observables to converge to thermal values do not immediately apply here.

\subsection{Convergence of the off-diagonal components}
\label{subsec:norm}

The initial state is given by a physical density operator, normalized in trace, $\mathrm{tr} \rho_0=1$, and also Frobenius norm, $\bra{\rho_0} \rho_0 \rangle= \mathrm{tr} \rho_0^2=1$. The filter \eqref{eq:Gaussianfilter} preserves the former, but not the latter. Instead, the norm of the filtered vector $\ket{\rho_{\sigma}}$ indicates the magnitude of the remaining off-diagonal components.

The state resulting from the application of the original Gaussian filter $F_{\sigma}$ on  
 $\rho_0$ can be written as a sum of two mutually orthogonal components, 
\beq
\ket{\rho_{\sigma}}
=\ket{\rho_{\mathrm{D}}}+\sum_{n,m\neq n} c_n c_m^{*} e^{-(E_n-E_m)^2/(2 \sigma^2)} \ket{E_n E_m}.
\eeq
The first term is precisely the diagonal ensemble, and the second one includes all off-diagonal components of the density operator.
Denoting them by $\ket{\Delta \rho}:= \ket{\rho_{\sigma}}-\ket{\rho_{\mathrm{D}}}$, the (Frobenius) norm of the off-diagonal components is
\beq
\bra{\Delta \rho} \Delta  \rho \rangle =\sum_{n,m\neq n} |c_n|^2 |c_m|^2 e^{-(E_n-E_m)^2/\sigma^2}.
\eeq

The magnitude of these components may be estimated using simple arguments. We consider as initial state $\rho_0$ a
 pure product state, for which the energy distribution, given by $|c_n|^2$, is peaked around the mean energy $E_{\rho_0}=\tr (H \rho_0)$, and has variance $O(N)$. For large systems, this distribution behaves as a Gaussian~\cite{Hartmann_2004} and
we can approximate the norm of the vector $\ket{\rho_{\sigma}}$ by a double integral over energies, from which we obtain

\beq
\bra{\rho_{\sigma}} \rho_{\sigma}\rangle \sim \frac{\sigma}{\sqrt{N}}.
\label{eq:FrobeniusNorm}
\eeq

The norm of the diagonal component, equivalent to the inverse participation ratio of the initial state, $\bra{ \rho_D}  \rho_D \rangle=\sum_n |c_n|^4$ is independent of $\sigma$.
Typically, the number of energy eigenstates contributing to the sum will be exponentially large in the system size, unless the mean energy of the initial state $E_{\rho}$ corresponds to a region of exponentially small density of states. To see this, we can take again into account the aforementioned distribution of the weights for our initial states, and the fact that for large systems the density of states approaches also a Gaussian distribution~\cite{Hartmann_2004,Keating_2015}. The inverse participation ratio then decreases exponentially with the system size,
\beq
\bra{\rho_{\mathrm{D}}} \rho_{\mathrm{D}}\rangle \sim 2^{-N}.
\eeq

Unless the width of the filter is exponentially small in $N$, the norm of the filtered state is dominated by the off-diagonal component, and we expect both of them to decrease proportionally to the width, for fixed size $N$, according to \eqref{eq:FrobeniusNorm}.
Notice nevertheless that a bound on the (Frobenius) norm of $\ket{\Delta \rho}$ is not enough to extract conclusions about the convergence of physical observables, a question that we explore numerically in section~\ref{Results}.

	\section{Setup for the numerical simulations}
	\label{Methods}
	
    We use numerical simulations to explore some of the questions in the previous section. In particular, we investigate whether 
	the diagonal ensemble can be approximated by a MPO,
	and how the physical observables approach the diagonal expectation values as we filter out the off-diagonal matrix elements of the density matrix.

\subsection{MPS approximation of the ensemble}

We use matrix product operators (MPO)~\cite{Verstraete-GarciaRipoll-Cirac_2004,ZwolakVidal2004,Pirvu_2010} to represent the density operators corresponding to the initial and filtered states. 
Once vectorized, they are represented by MPS with double physical indices, which can be manipulated using standard tensor network methods~\cite{Verstraete2008,SCHOLLWOCK2011,ORUS2014117,Silvi2019}.

We find a MPS approximation for the action of the filter \eqref{eq:chebyfilter} on a given initial state.
 The method is completely analogous to the one presented in~\cite{banuls2019entanglement} for filtering out energy fluctuations, with the only difference that here the effective Hamiltonian is the commutator superoperator $\Hc$ acting on the vectorized density matrices. For a local Hamiltonian $H$, the commutator $\Hc$ can also be written as a MPO.
 
 As in~\cite{Holzner2011,Halimeh2015,WolfSpect2015,Xie2018,banuls2019entanglement,Yang_2020} 
	we can then take advantage of the fact that we do not need the full polynomials $T_m({\Hc})$, which in our case are operators acting on a $2^{2N}$ dimensional vector space, but only the vectors resulting from their action on the initial state $T_m({\Hc})|\rho_0\rangle$. The latter satisfy the same recurrence relation as the polynomials and
	can be computed with lower computational cost.

	\subsection{Model and initial states}
We focus our study in the non-integrable Ising spin chain with longitudinal and transverse fields,
	\beq
	H_{\rm Ising}=J\sum_{i} \sigma_z^{[i]}\sigma_z^{[i+1]}+g\sum_{i} \sigma_x^{[i]}+h\sum_{i} \sigma_z^{[i]},
	\label{eq:model}
	\eeq
and choose parameters $(J,g,h)=(1,-1.05,0.5)$, which is far from the integrability limit.

As initial  states we consider product states in which all spins are aligned in the same direction. We denote such states by the direction in which the spins are aligned, e.g. $\ket{X\pm}=2^{-N/2}\left (\ket{0}\pm\ket{1}\right)^{\otimes N}$, $\ket{Y\pm}=2^{-N/2}\left (\ket{0}\pm i\ket{1}\right)^{\otimes N}$, $\ket{Z+}= \ket{0}^{\otimes N}$
and $\ket{Z-}= \ket{1}^{\otimes N}$.
	
	\section{ Numerical Results}\label{Results}

	We have applied the procedure described in the previous section to system sizes  $N\in\{20,\,60\}$, using MPS with bond dimensions $100\leq D \leq 1500$. Additionally, we cross-check results for small system sizes $N\leq 20$ which can be explored with exact diagonalization.

	\subsection{Scaling}
	\label{subsec:scaling}
	
	\begin{figure}[h!]
	\begin{center}
    \includegraphics[width=.95\columnwidth]{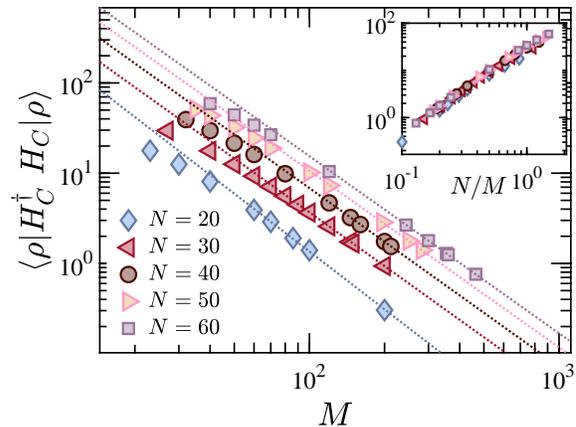} 
	\caption{Scaling of the variance $\delta^2= \langle \rho_M |\Hc^{\dagger} \Hc|\rho_M \rangle$, as a function of the Chebyshev truncation parameter $M$ for different system sizes $N=20-60$ with bond dimension $D=1000$ and initial state $\ket{X+}$. 
	Except for the smallest values of $M$, we find that our results scale with the expected~\cite{banuls2019entanglement} $\delta^2\propto 1/M^2$.}	\label{Fig1:scaling}
	\end{center}
    \end{figure}

We expect the off-diagonal width $\delta$ of our simulations to follow the scaling predicted in Ref.~\cite{banuls2019entanglement}, namely $\delta^2 \propto 1/M^2$, for large enough number of terms in the approximation of the filter, and provided that the truncation error is not significant.
Thus, the decrease of the width with $M$ provides us with a check that our simulations are in the expected regime.
Figure~\ref{Fig1:scaling} shows that
this is indeed the case.
The figure shows that for all system sizes, the converged data are well described by a power law fit $\delta^2\propto M^{-\alpha}$ (dotted lines) with exponents $-2.13, -1.98, -1.97, -1.95, -1.96$ for $N=20, 30, 40, 50, 60$, respectively.

A further check is provided by the
 norm of the filtered state $\ket{\rho_{\sigma}}$. 
As described in section~\ref{subsec:norm},
$\bra{\rho_{\sigma}}\rho_{\sigma}\rangle$ should decrease as the inverse off-diagonal width. Since our algorithm applies the normalized filter \eqref{eq:chebyfilter}, $Q_M \sim \frac{1}{\sqrt{2 \pi \sigma^2}} F_{\sigma}$, we expect, 
 for the proper values of $M$ and $\sigma$, 
\beq
\bra{\rho_M} \rho_M \rangle \sim \frac{1}{\sigma\sqrt{N}}.
\eeq

\begin{figure}[h!]
    \begin{center}
    \includegraphics[width=.95\columnwidth]{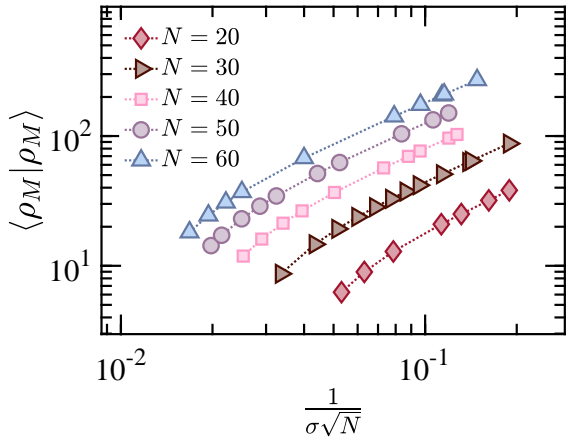} 
	\caption{Relation between vector norm of off-diagonal components and inverse off-diagonal width for system sizes $N=20-60$ and bond dimension, $D=1000$, starting with initial state $\ket{X+}$. }	\label{Fig2:normMPS}
	\end{center}
	\end{figure}

To directly probe this relation, we plot the vector
norm of our resulting state in figure~\ref{Fig2:normMPS}, for system sizes $N=20-60$, and find that our data agrees well with this prediction, except for the smallest values of $M$.

	\subsection{Convergence of local observables}
	\label{subsec:observables}
	\begin{figure}[h!]
	\begin{center}
	\includegraphics[width=.96\columnwidth]{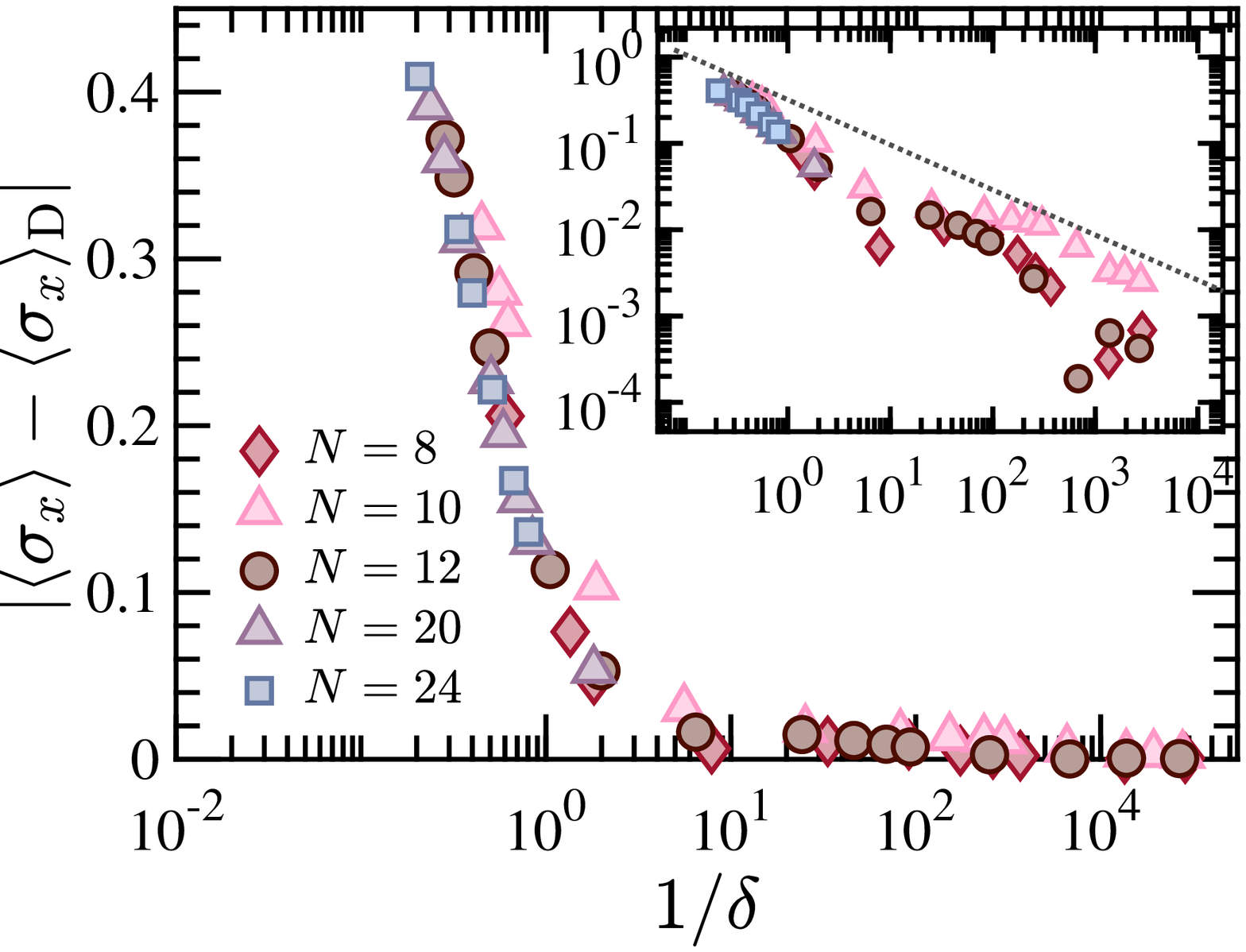}
	\includegraphics[width=.96\columnwidth]{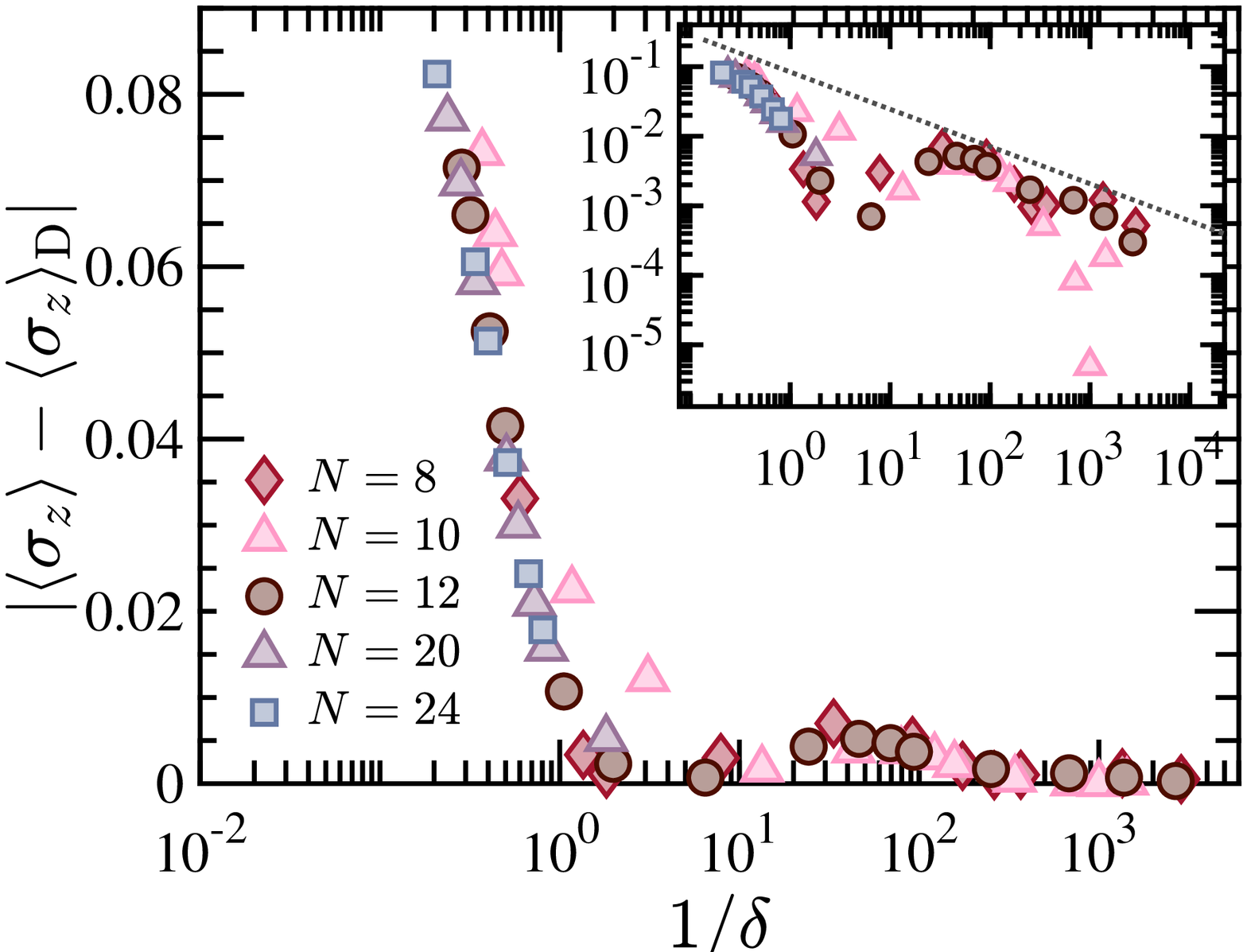}
	\caption{Absolute error in local observables $\sigma_x$ (upper) and $\sigma_z$ (lower figure) between exact diagonal ensemble values and Chebyshev filter results as a function of inverse off-diagonal width for system sizes $N=8, 10, 12, 20$ and $24$ with the initial state $\ket{X+}$. The insets indicate the log-log plot of the corresponding figures, which we show the upper bounds with the straight dotted lines. The slope for $\sigma_x$ is -0.52 and it is -0.53 for $\sigma_z$.}	\label{Fig3:AbsErrSmallSys_initX}
	\end{center}
	\end{figure}
	
	\begin{figure}[h!]
	\begin{center}
	\includegraphics[width=.96\columnwidth]{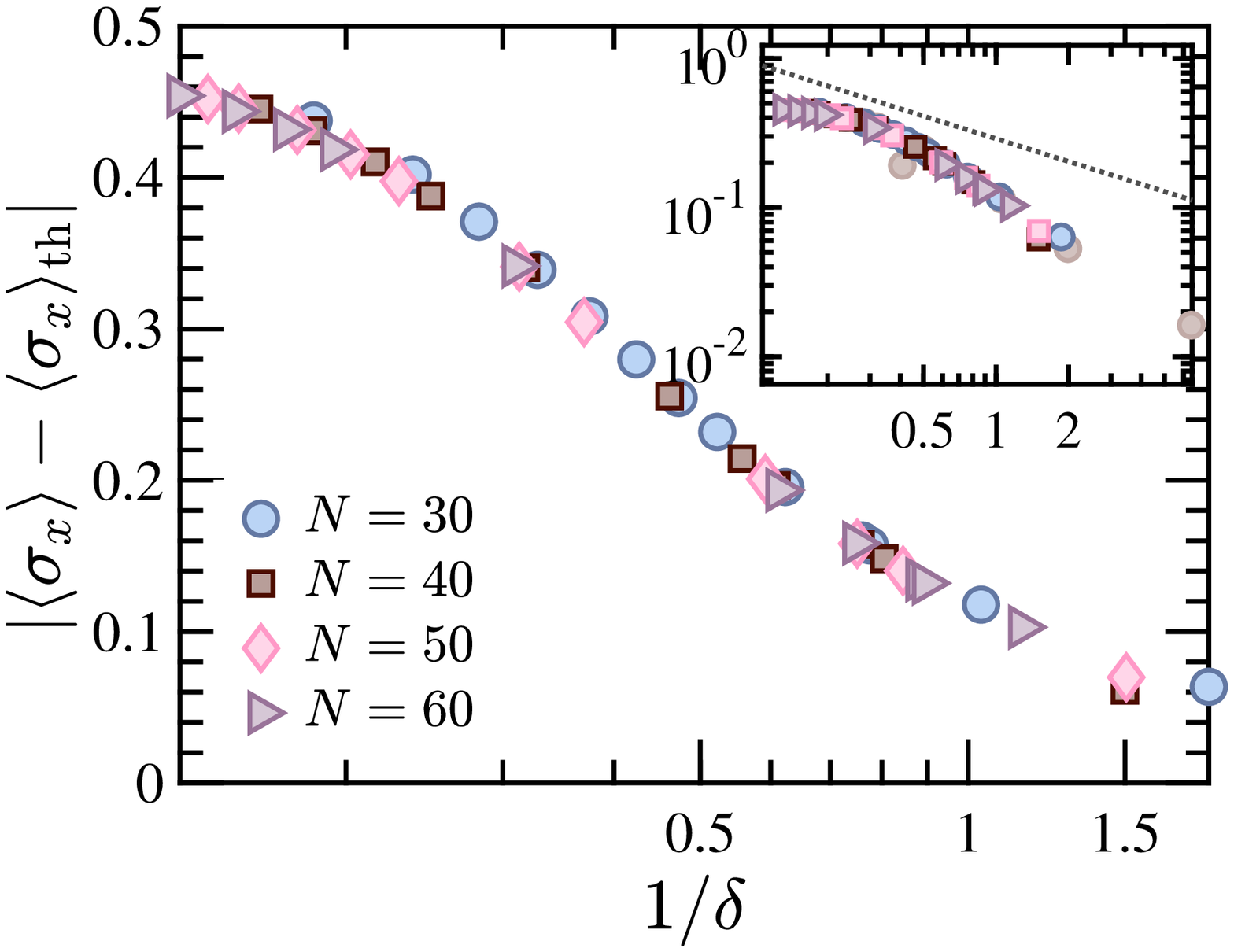}
	\includegraphics[width=.96\columnwidth]{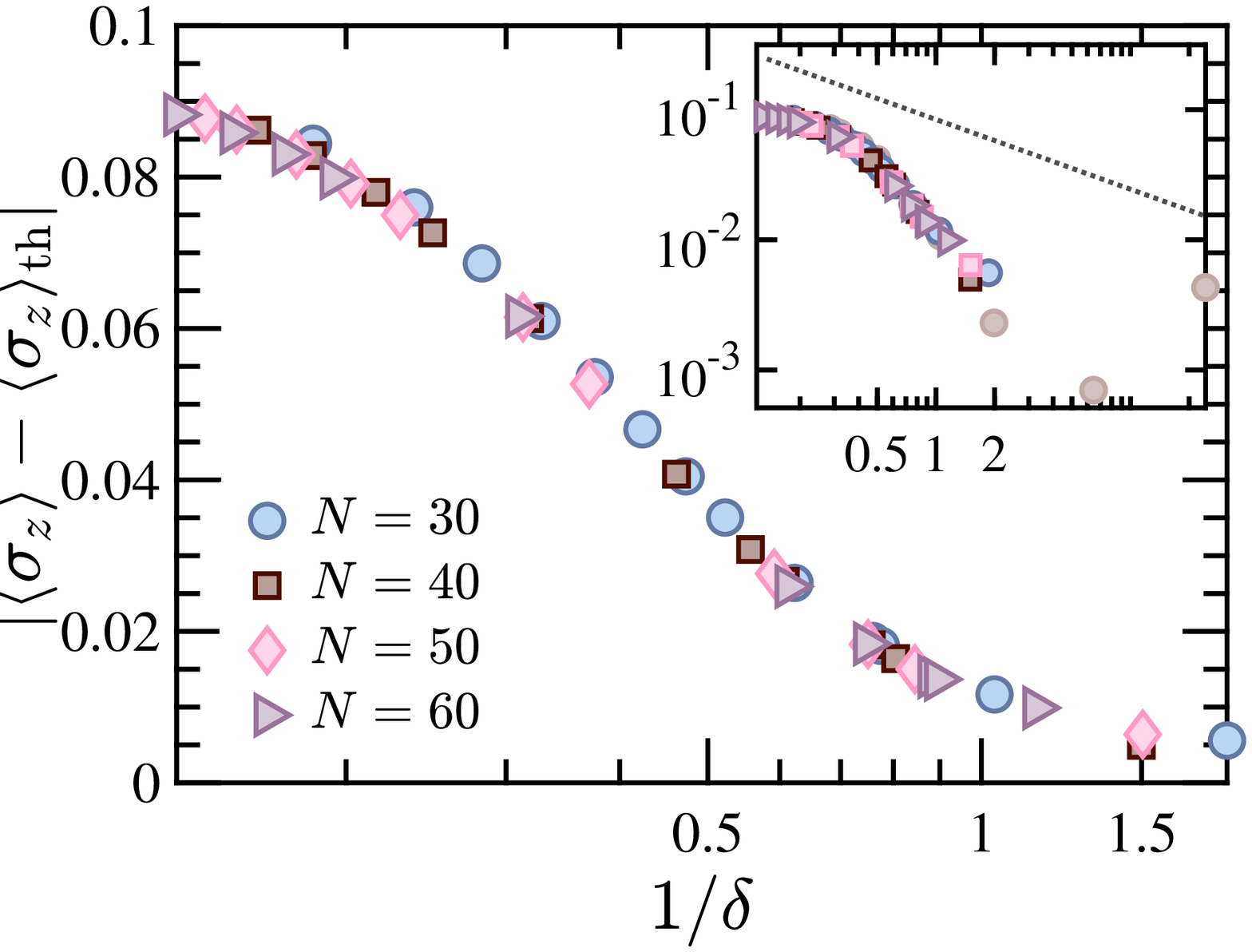}
	\caption{Absolute error in local observables $\sigma_x$ (upper) and $\sigma_z$ (lower figure) between thermal values and numerical results based on Chebyshev filter as a function of inverse off-diagonal width for system sizes $N=30-60$ with the initial state $\ket{X+}$. The insets indicate the log-log plot of the corresponding figures, where we add the upper bounds with the dotted lines and the data points belong to $N=12$ with lighter color as reference values taken from figure~\ref{Fig3:AbsErrSmallSys_initX}.}	\label{Fig4:AbsErrLargeSys_initX}
    \end{center}
	\end{figure}	
	As the filtered state approaches the diagonal ensemble, so will the values of physical observables.
	If the state thermalizes, such limit will agree with the thermal value corresponding to the initial energy, and thus comparing this to the converged values can be used to probe thermalization of the system. Here we are interested in the rate of convergence of the physical expectation values.

For the problem of reducing the energy variance of a pure state,
it has predicted that for chaotic systems~\cite{Dymarsky2019} a polynomial decrease of the variance with the system size is required for all local observables to 
converge to their thermal values. 
In Ref.~\cite{banuls2019entanglement} it was numerically observed for model \eqref{eq:model} that
an energy variance decreasing as $1/\log N$ or faster was sufficient for convergence in the thermodynamic limit. But as discussed in section~\ref{Filtering}, these conclusions do not need to apply in our case, because the expectation value in the mixed state does not have the same local structure. We thus explore this question numerically by studying the local $x$ and $z$ magnetizations in the middle of the chain, $\mathcal{O}=\sigma_{x,z}^{[N/2]}$, and analyzing how the expectation values vary as the width of the filter decreases.
For systems of size $N\leq 12$ we can compute the action of the filter exactly for any width, while for larger systems, up to $N\leq 60$, we run MPS simulations up to the narrowest filter widths that we can reliably reach with a maximum bond dimension $D=1000$.

For small systems, $N\leq 24$, we can compare the filtered values to the exact  compute the exact magnetizations in the diagonal ensemble. 
For larger systems we do not have access to either the evolved state at long times or the exact diagonal ensemble, but we can approximate the thermal ensemble corresponding to the initial energy using MPO~\cite{Verstraete-GarciaRipoll-Cirac_2004,ZwolakVidal2004,Feiguin-White-2005}. For the cases we study, there are analytical and numerical arguments in favor of thermalization~\cite{Lin2019,Yang_2020}, such that the thermal expectation values should be very close to the diagonal ones. Thus, for our analysis it is enough to use the thermal value as reference, since we are only exploring the variation of the expectation values, but our simulations for large systems do not reach full convergence (see subsection~\ref{Error Analysis} for a more detailed discussion of the numerical errors).

\begin{figure}[h!]
	\begin{center}
	\includegraphics[width=.95\columnwidth]{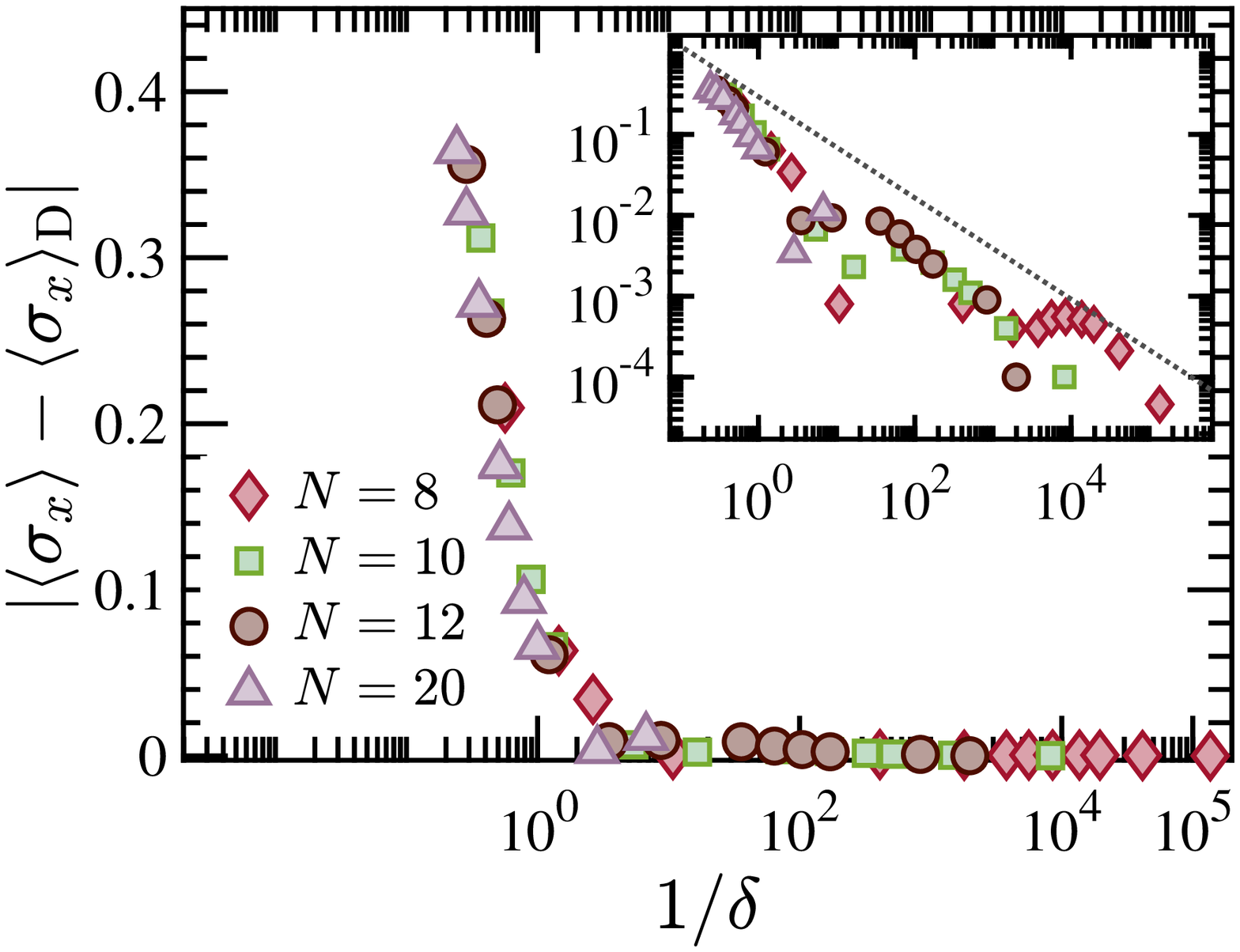}
	\includegraphics[width=.95\columnwidth]{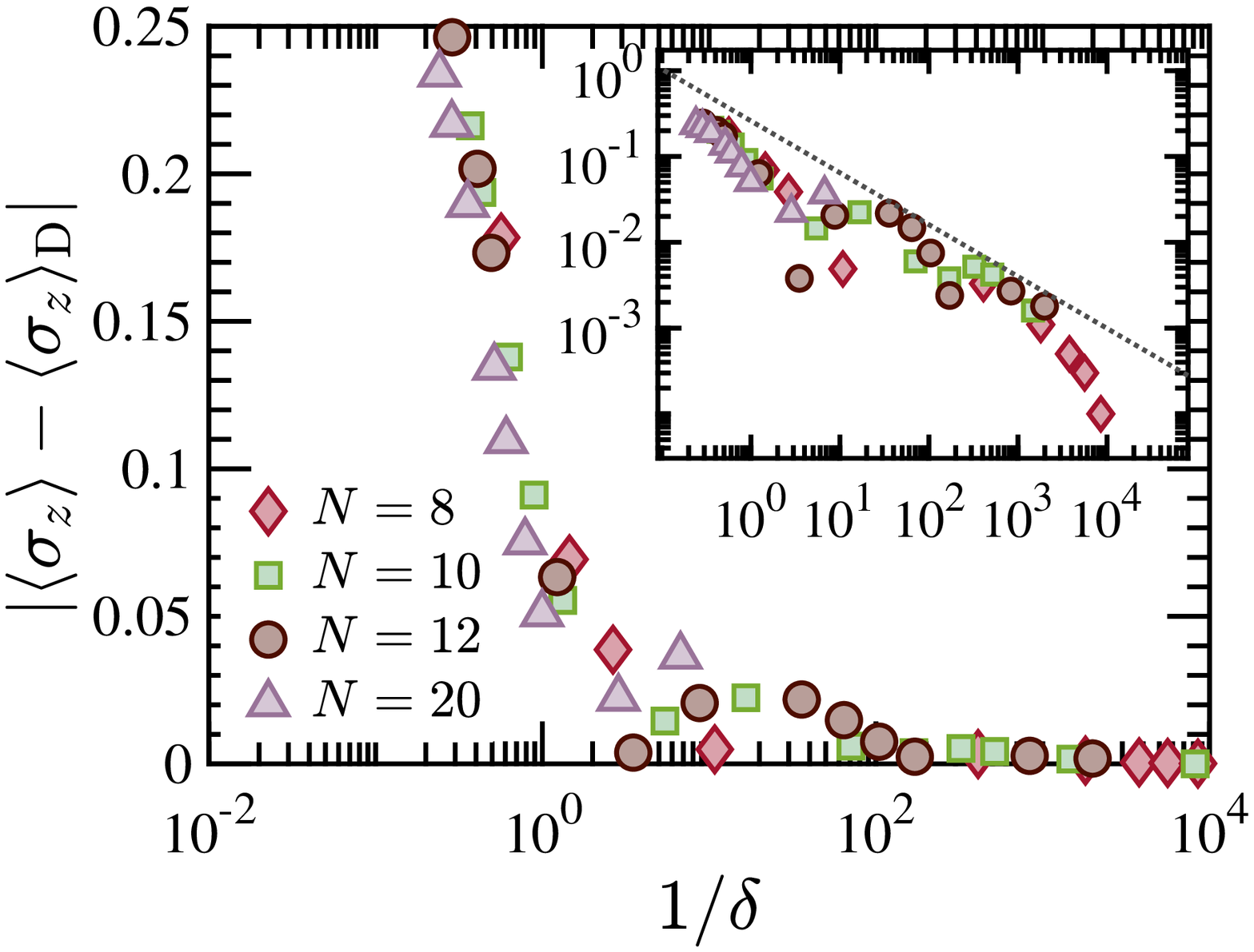}
	\caption{Absolute error in local observables $\sigma_x$ (upper) and $\sigma_z$ (lower figure) between exact diagonal ensemble values and Chebyshev filter results, as a function of inverse off-diagonal width for system sizes $N=8, 10, 12$ and $20$ with the initial state $\ket{Z+}$. The insets indicate the log-log plot of the corresponding figures, which we show the upper bounds with the grey  dotted lines. Slopes for $\sigma_x$ is -0.62(63) and it is -0.60(47) for $\sigma_z$.}
	\label{Fig5:AbsErrSmallSys_initZ}	\end{center}
	\end{figure}

\begin{figure}[h!]
	\begin{center}
	\includegraphics[width=.94\columnwidth]{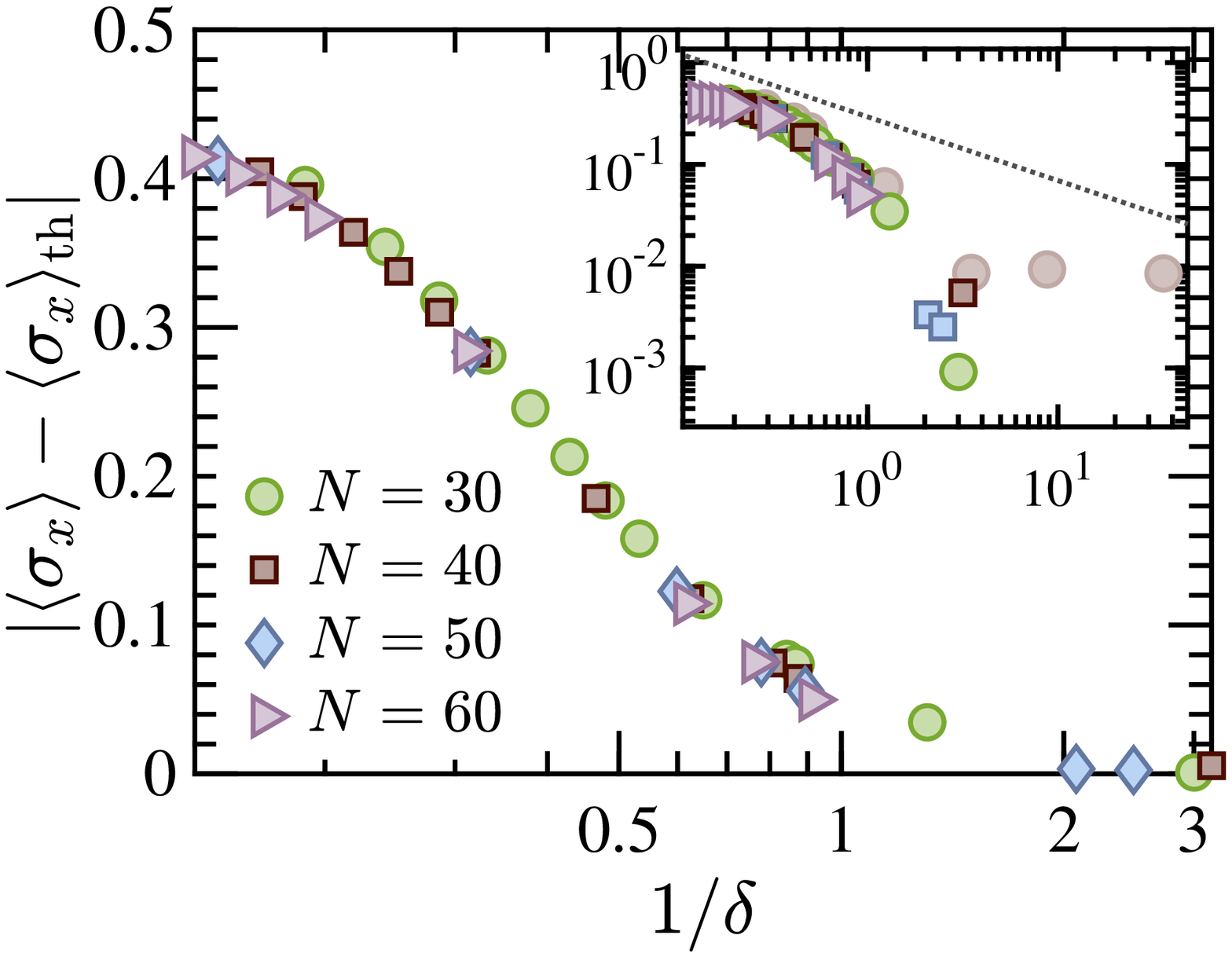}
	\includegraphics[width=.94\columnwidth]{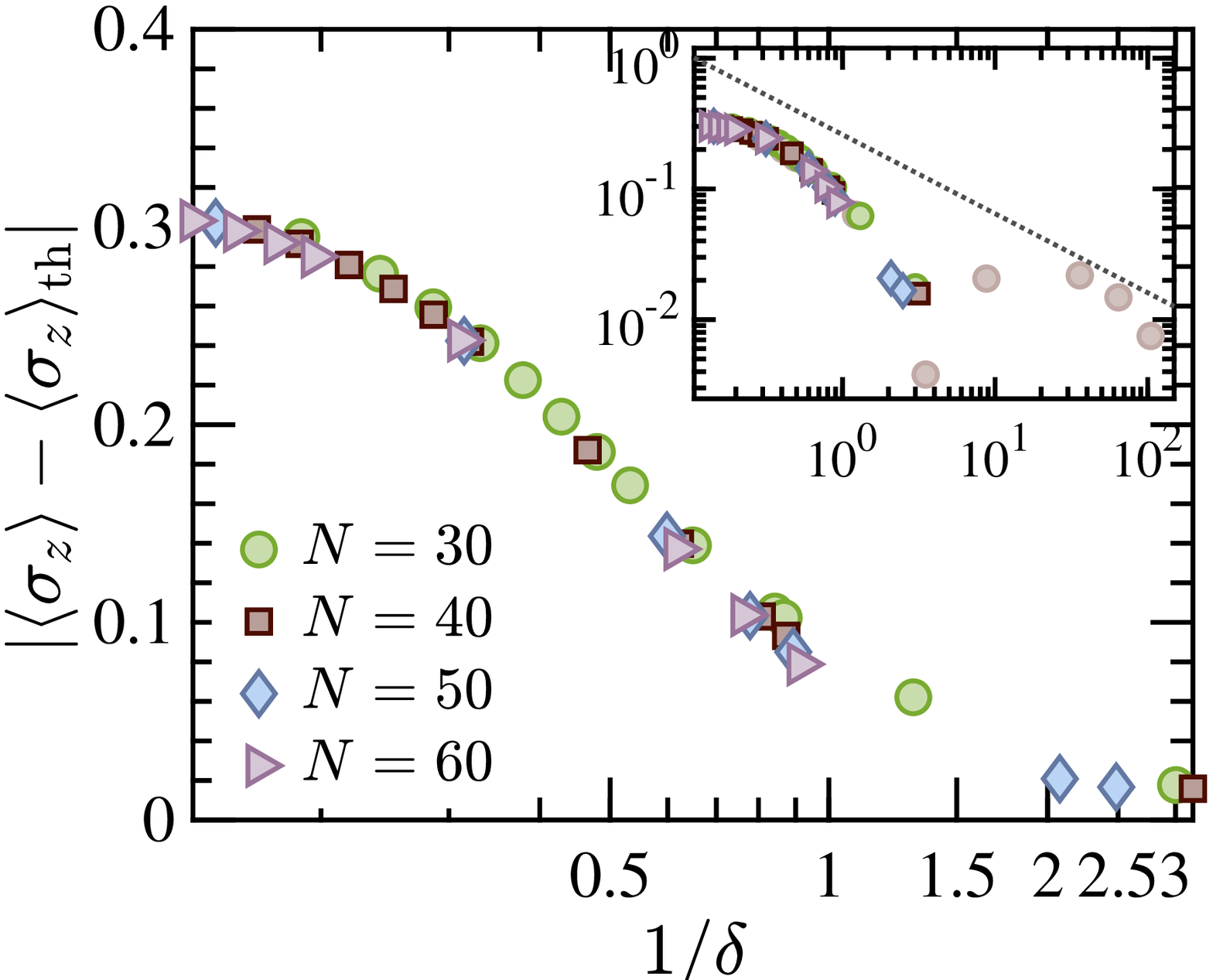}
	\caption{Absolute error in local observables $\sigma_x$ (upper) and $\sigma_z$ (lower figure) between thermal values and numerical results based on Chebyshev filter as a function of inverse off-diagonal width for system sizes $N=30-60$ with the initial state $\ket{X+}$. The insets indicate the log-log plot of the corresponding figures, where we put the upper bounds with the dotted lines and the data points belong to $N=12$ with lighter color as reference values taken from figure~\ref{Fig5:AbsErrSmallSys_initZ}.}	\label{Fig6:AbsErrLargeSys_initZ}
    \end{center}				
	\end{figure}

We plot the results for small and large system sizes in figures~\ref{Fig3:AbsErrSmallSys_initX} and~\ref{Fig4:AbsErrLargeSys_initX}
for initial state $\ket{X+}$, 
and in figures~\ref{Fig5:AbsErrSmallSys_initZ} and~\ref{Fig6:AbsErrLargeSys_initZ} for initial state $\ket{Z+}$.
In all cases we represent the absolute value of the difference between the expectation values in $\ket{\rho_M}$ and the diagonal (thermal, for large systems) values as a function of the off-diagonal width $\delta$.
In all cases, i.e. for the different initial states and different sizes, we observe that this absolute error, which is given exclusively by the off-diagonal part of $\rho_M$, decreases
at least as fast as $1/\sqrt{\delta}$ (see insets).
Moreover, the figures show that curves for different system sizes practically collapse on top of each other.

	\subsection{Entropy}
	\label{subsec:entropy}
	Since we start with a product state $\ket{\rho_0}$ and evolve it with a local Hamiltonian $\Hc$, the same arguments used in the case of pure states~\cite{banuls2019entanglement,Acoleyen_2013} then
	imply that the OSEE can be bounded as a function of the off-diagonal width and the system size as given in Eq.~\eqref{eq:ent}.

    \begin{figure}[ht]
    \begin{center}
	\includegraphics[width=.92\columnwidth]{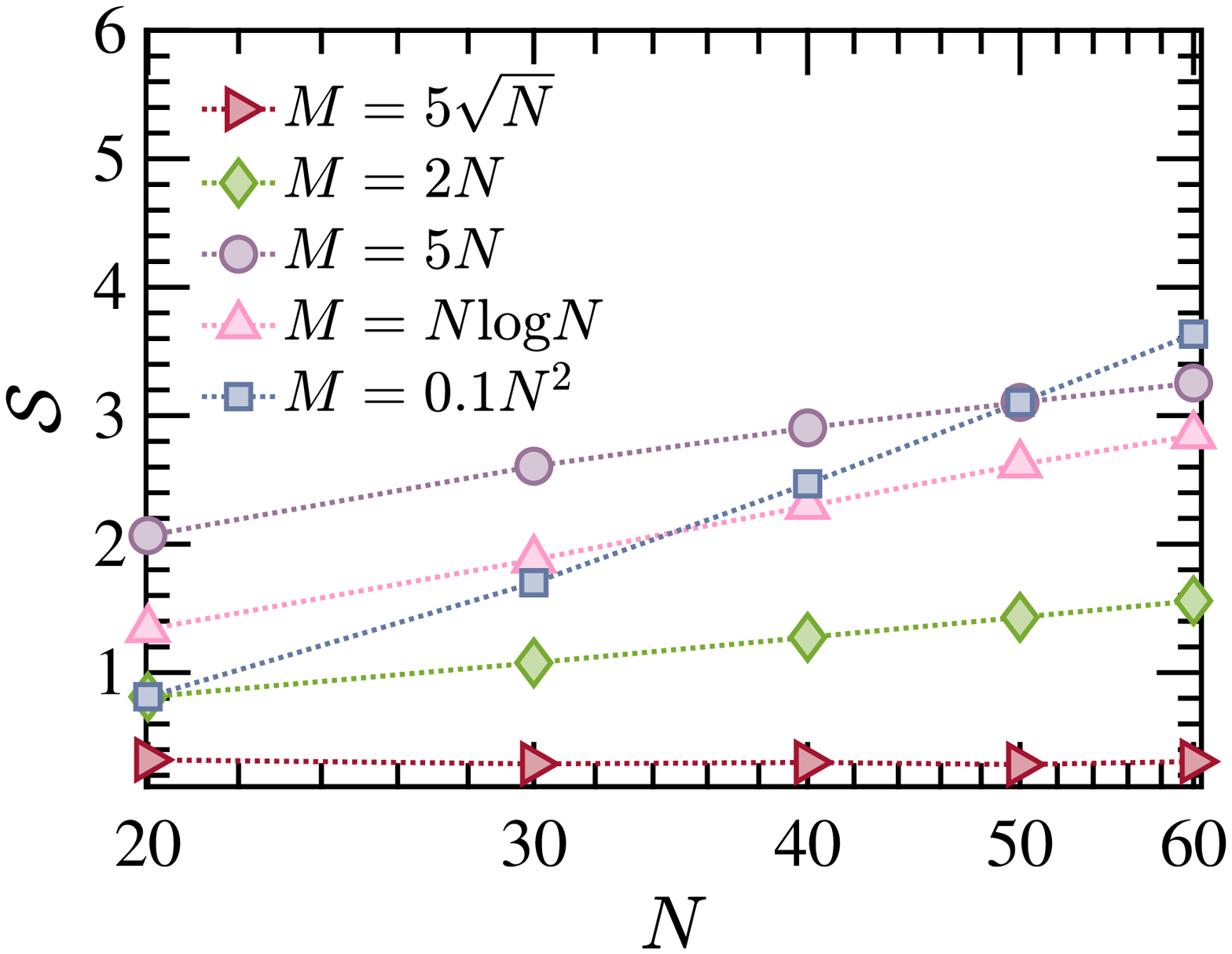}
	\includegraphics[width=.92\columnwidth]{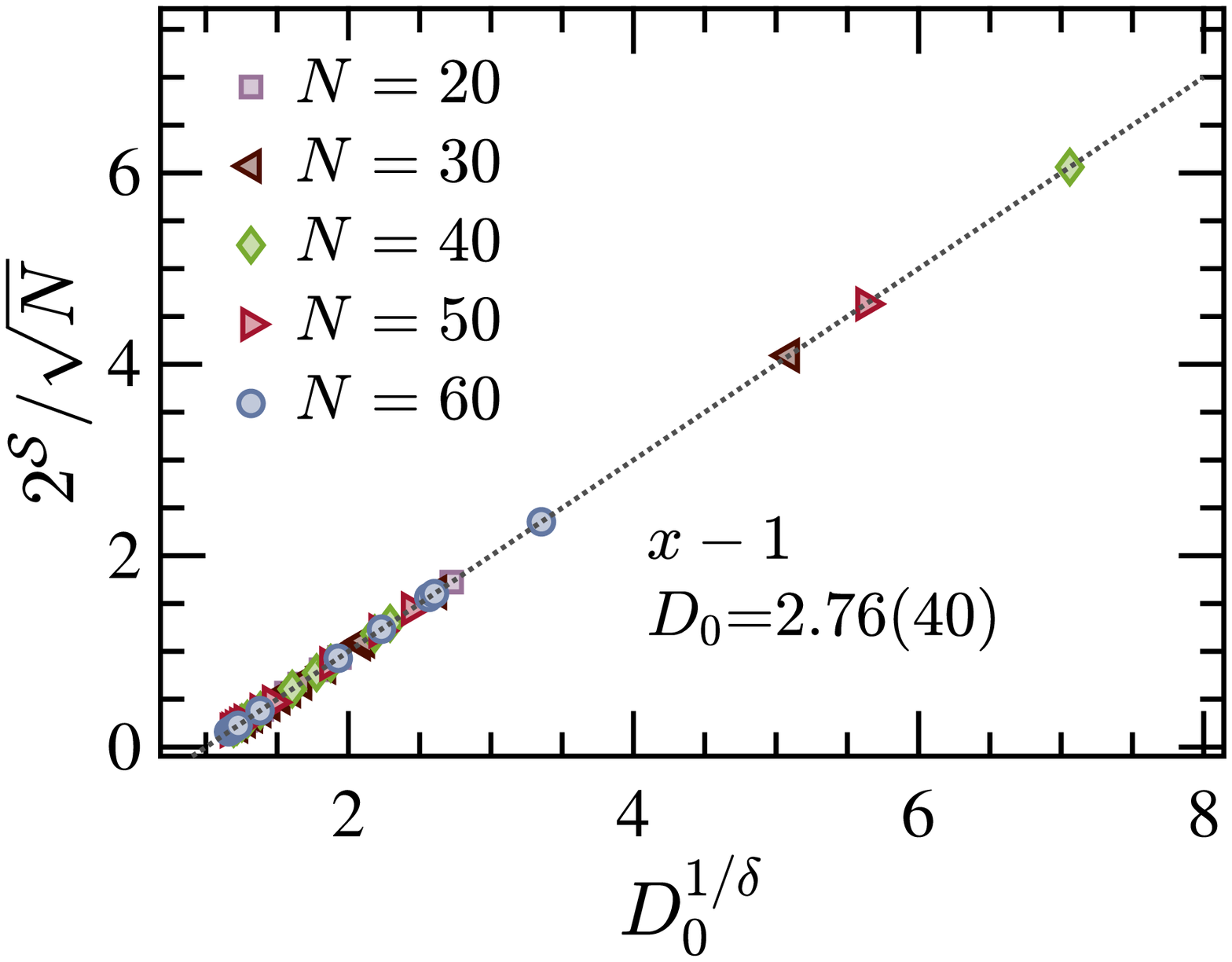}
	\caption{Left figure: Operator space entanglement entropy of the half chain as a function of logarithm of the system size, $N$, with different truncation numbers of Chebyshev filter, $M=f(N)$ and bond dimension, D=1000 for initial state $\ket{X+}$. Our data show that the entropy grows with $\log N$ in all cases except that the line for $M=5\sqrt{N}$ stays constant. Right figure: Behavior of the exponential of the entropy as predicted by Ref.~\cite{banuls2019entanglement} that we have shown in eq.~\ref{eq:ent}. The dotted line indicates the linear fit where all data points locate on the same line as expected for large system sizes. $D_0$ from fitting the data for all system size is $2.76(40)$ and the slope of the fit is $1$.}	\label{Fig7:SVsVar_largeSys}
    \end{center}	
    \end{figure}

	Figure~\ref{Fig7:SVsVar_largeSys} (upper) shows that indeed, the evolution of the OSEE while filtering out the off-diagonal components of the state satisfies a similar bound.
	The plot shows the OSEE corresponding to the middle cut of the approximate filtered state $\rho_M$, as a function of the system size, for simulations in which the number of Chebyshev terms was chosen as different functions of the size $M=f(N)$, corresponding to a width $\delta(N) \propto 1/M$.
	We observe that for $M\propto\sqrt{N}$, which corresponds to $\delta \propto\sqrt{N}$, the OSEE does not grow with the system size, while for $M\propto N$ or $M\propto N \log N$ (correspondingly $\delta \sim \mathrm{const}$ or $\delta \propto 1/\log N$ ), it increases as $\log N$.
	For faster growing $M\propto N^2$, also the increase in entropy is faster (compatible with it growing at most as $N$, as predicted by the argument in \cite{banuls2019entanglement}).

The asymptotic universal scaling of the entropy can be appreciated more explicitly in figure~\ref{Fig7:SVsVar_largeSys} (lower), which shows that $2^{\mathcal{S}}\propto \sqrt{N}(D_0^{1/\delta}-1)$ for all system sizes $N\geq 20$ with a constant $D_0=2.76$.

    \begin{figure}[htbp]
    \hspace{-0.5cm}   \includegraphics[width=.54\columnwidth]{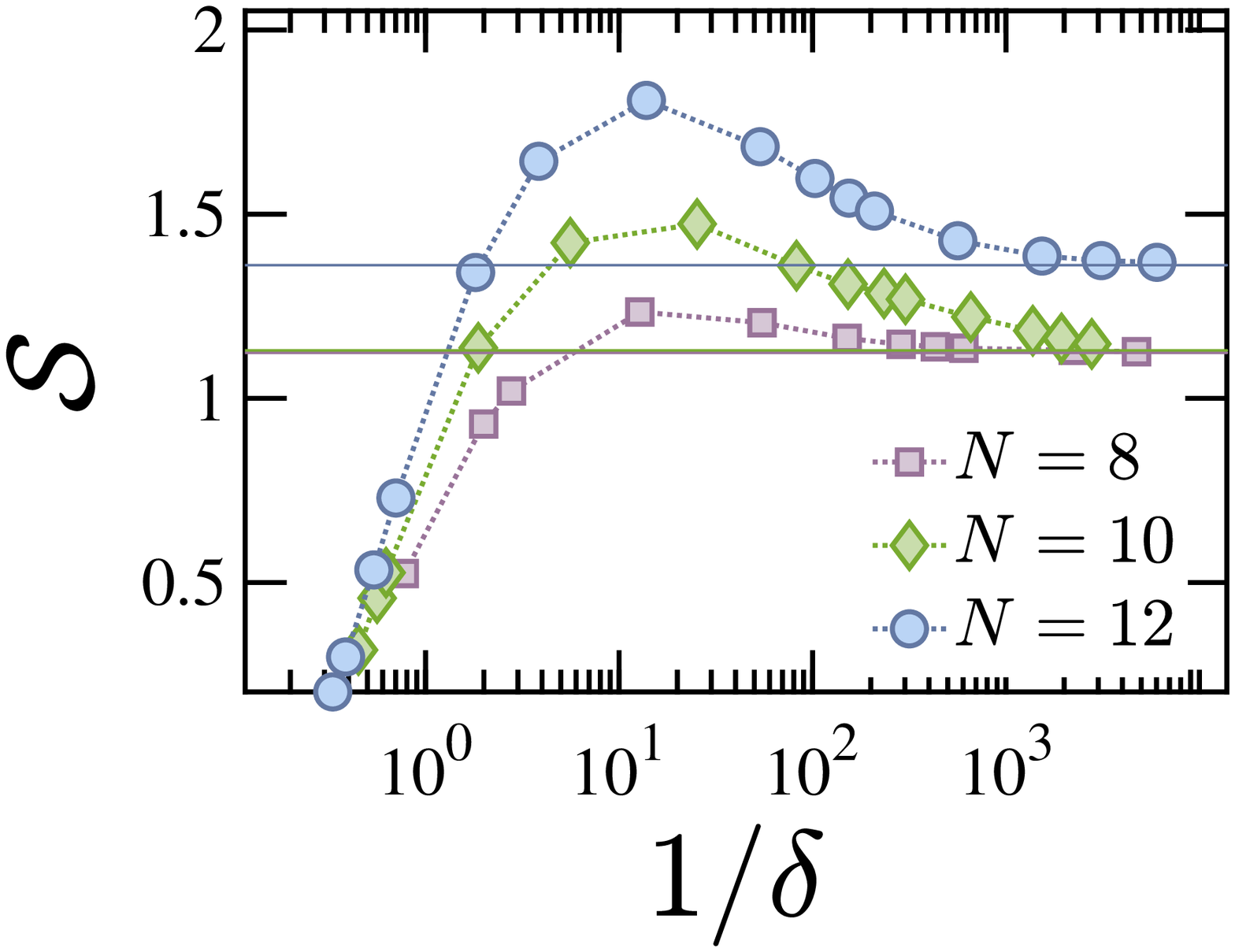}\hspace{-0.52cm}
	\includegraphics[width=.54\columnwidth]{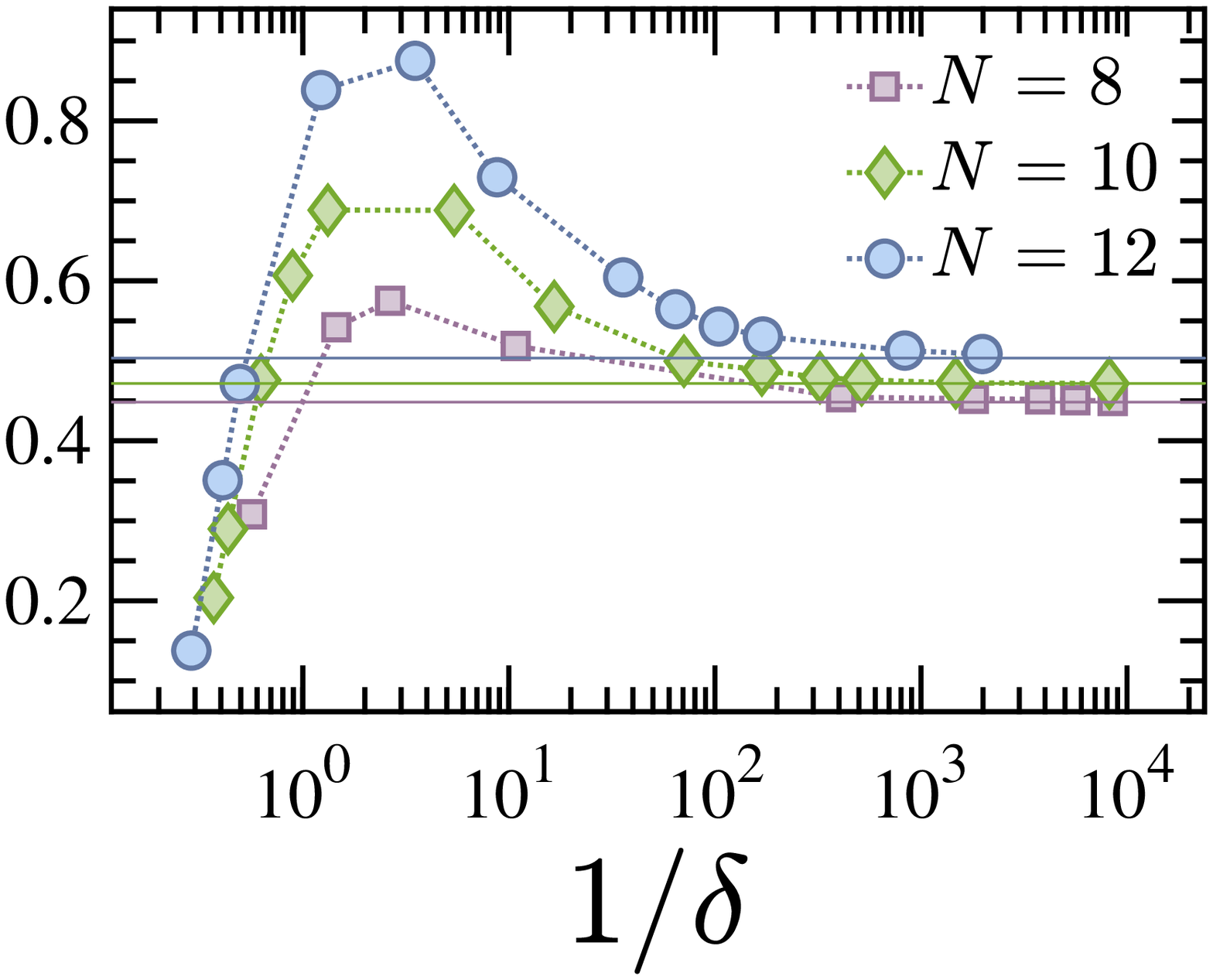}
	\caption{Relation between entropy and log of $1/\delta$ based on exact calculation for $N=8, 10, 12$ with initial state, $\ket{X+}$ (left) and $\ket{Z+}$ (right figure).}	\label{Fig8:exactSVsVar}
    \end{figure}

The limit of the filtering procedure when the width vanishes is a mixed state in the exponentially degenerate null space of $\Hc$. This subspace supports states with zero OSEE (e.g. the maximally mixed state), and thus the final OSEE is not generic, but will be determined by the initial state, in contrast to the case of pure state filtering, where we could generically expect that the entanglement entropy converged to a thermal volume law. We can explore how the limit value is approached during the filtering by analyzing the results for small systems, as shown in figure~\ref{Fig8:exactSVsVar}. As illustrated in the figure for different initial states and sizes $N\leq 12$, the entropy grows with $1/\delta$ for moderate widths, but it reaches a maximum after a certain point, and then decreases towards the diagonal value. If we examine how this final value depends on the system size, we observe, that in all the cases studied the diagonal OSEE increases almost linearly with the size, although the values change considerably from one state to another, where the slope of each initial states are $0.06(78), 0.89(17), 0.02(37)$ for $\ket{X+}, \ket{Y+}, \ket{Z+}$, respectively.

\subsection{Error Analysis} \label{Error Analysis}
	
\begin{figure}[h!] 
	\begin{center}
	\includegraphics[width=.52\columnwidth]{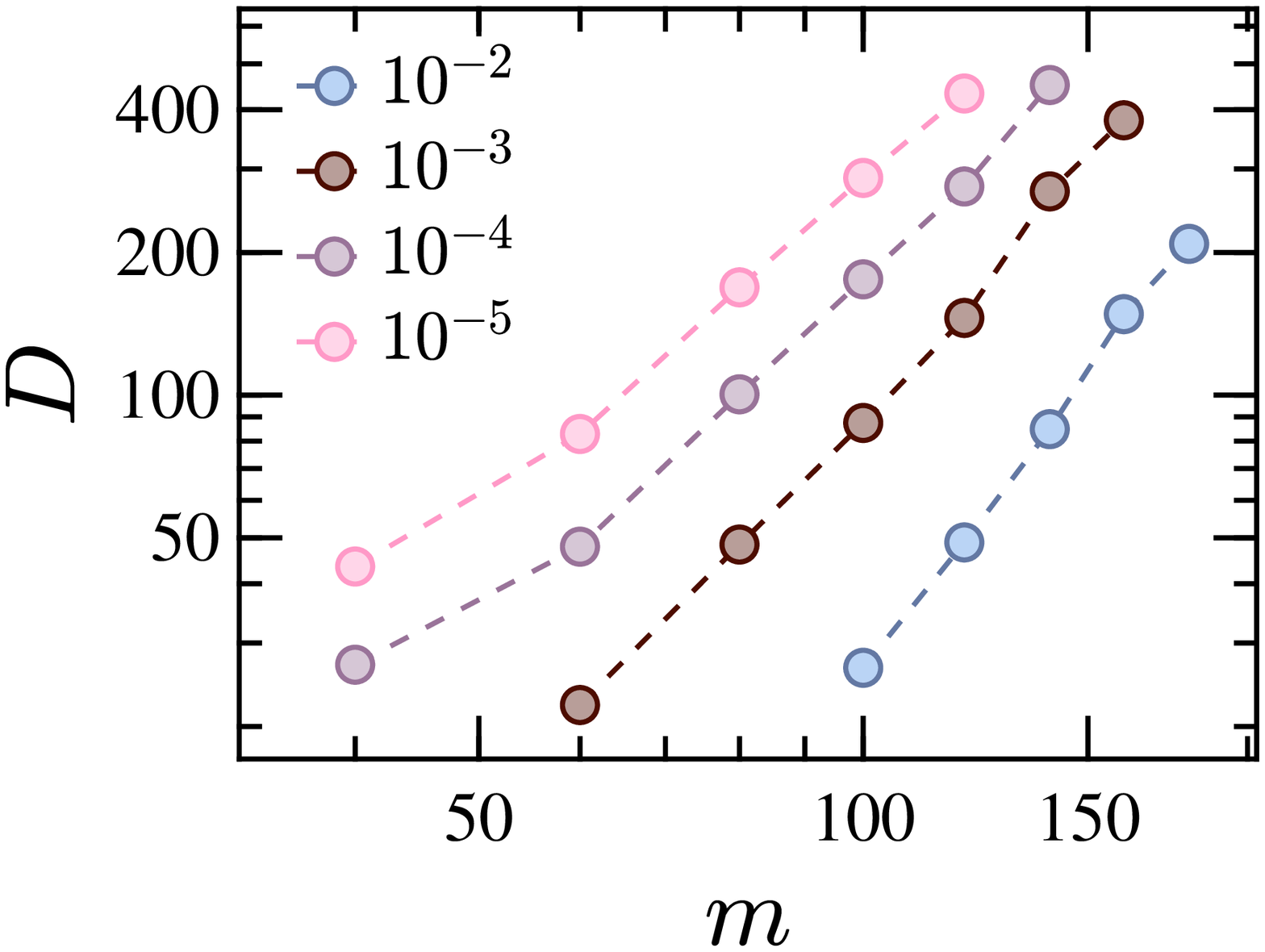} 	\hspace{-0.55cm}
	\includegraphics[width=.52\columnwidth]{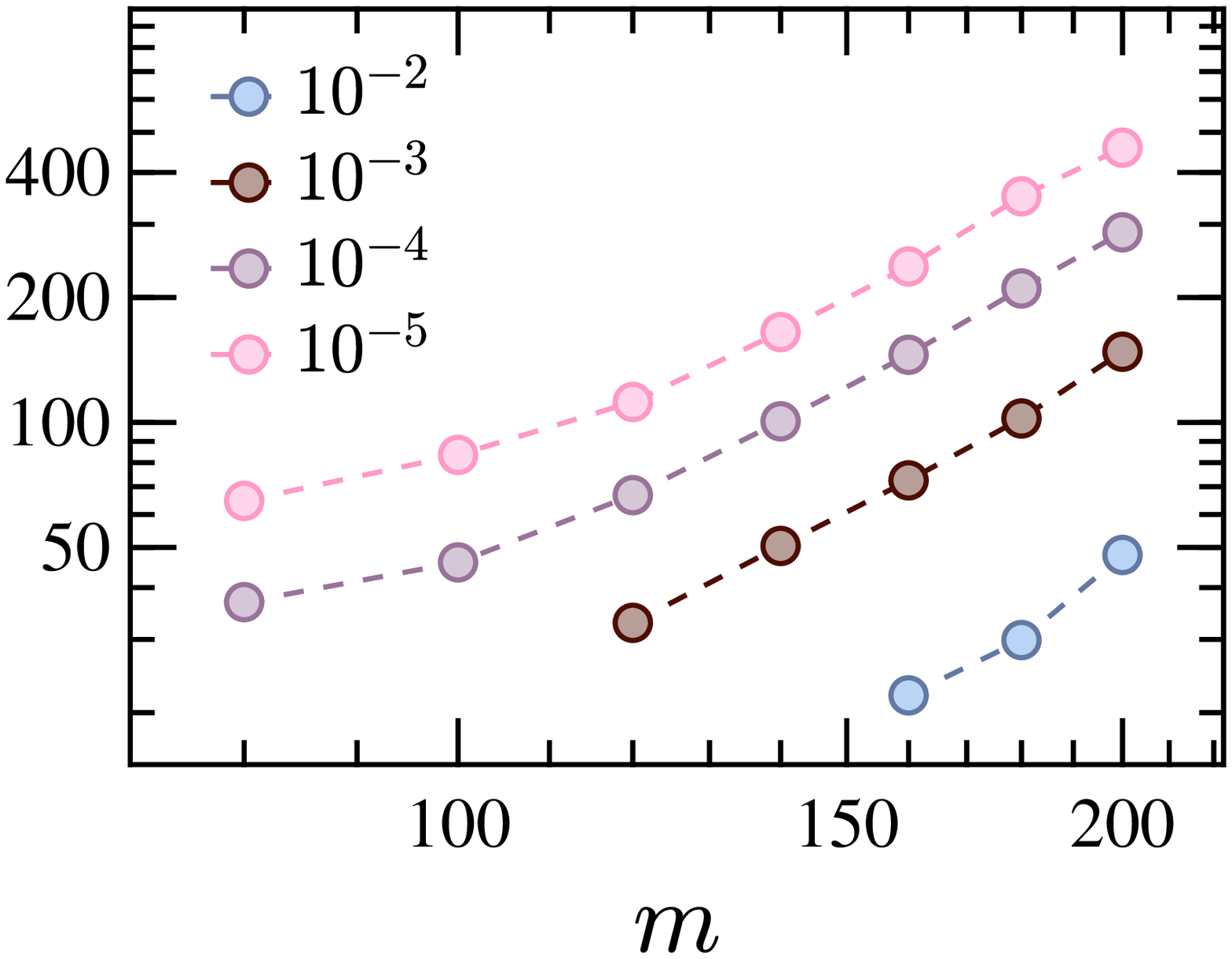} 	
	\caption{Scaling of the bond dimension required to keep a constant precision in the MPS approximation of $T_m(\Hc)\ket{\rho_0}$, as a function of the degree $m$ for various values of the truncation error, $10^{-2}, 10^{-3}, 10^{-4}, 10^{-5}$ and system sizes $N=20$ (left) and $N=30$ (right) for $D=500$.} \label{requiredBondDim}
	\end{center}
	\end{figure}

In our strategy, for a fixed order $M$ of the Chebyshev expansion, the main source of error is the truncation error, namely approximating the action of each Chebyshev polynomial on the initial state by a MPS with limited bond dimension. We can quantify this error for a given order $m$ using as reference the best approximation found for the corresponding term $T_m(\Hc)\ket{\rho_0}$ (in our case, with $D=1000$) and comparing it to its truncated versions with smaller bond dimensions. In this way we can extract the bond dimension required for fixed precision.
In previous works that used MPS approximations of Chebyshev series~\cite{Holzner2011,Halimeh2015,WolfSpect2015,Xie2018,Yang_2020} it was observed that the required bond dimension for such terms increases polynomially with the degree $m$. 
Our results, illustrated in figure~\ref{requiredBondDim}, seem to agree with such behavior, except for the smallest values of $m$. 
We have also observed, as in the recent work~\cite{Yang_2020}, that for fixed $m$ the bond dimension required to maintain constant truncation error in $T_m(\Hc)\ket{\rho_0}$ gets smaller for larger system sizes.
Notice, however, that for larger systems, also polynomials of higher degree will be required to attain a constant width $\delta$, since, as discussed in Sect.~\ref{Chebyshev approximation of the filter}, the order of the expansion scales as $M\propto N/\delta$.

\section{Discussion}
\label{Discussion}

We have presented a method to approximate the diagonal ensemble corresponding to a quantum many-body state. 
By applying a Gaussian filter to the density operator, the off-diagonal components in the energy basis are suppressed and, in the limit of vanishing filter width, the result converges to the ensemble that represents the long time average of the time evolved state. For a Hamiltonian with non-degenerate spectrum, this is the diagonal ensemble.

Numerically, the filter can be approximated by a Chebyshev polynomial series, and applied using MPS standard techniques, in an analogous manner to what was already described in Ref.~\cite{banuls2019entanglement} for an energy filter. In our case, we obtain a MPO approximation to the filtered ensemble. 

The method allows us to treat larger systems than exact diagonalization. However our results for small systems indicate that the operator space entanglement entropy of the diagonal ensemble scales as a volume law, which limits the system sizes for which the MPO can provide a reliable approximation. Still, we are able to simulate the effect of filters with moderate off-diagonal width and to analyze the convergence of local observables towards the thermal equilibrium. 

We have applied this method to a non-integrable spin chain and several out of equilibrium product initial states for system sizes up to $N=60$. We have numerically observed that local observables converge towards their thermal values as a power of the inverse off-diagonal width.
Remarkably, this behavior is mostly independent from the system size.
Even for moderate off-diagonal widths, the method provides in this way insight beyond  exact diagonalization. In the future, it can be thus used to explore other one-dimensional models.

\acknowledgments

This work was partially funded by the Deutsche Forschungsgemeinschaft (DFG, German Research Foundation) under Germany's Excellence Strategy -- EXC-2111 -- 390814868 and by the European Union through the ERC grant QUENOCOBA, ERC-2016-ADG (Grant no. 742102).

%\bibliography{diagonalEnsemble} 
\input{diagonalEnsemble.bbl}
\end{document}